\documentclass[fleqn,10pt]{wlscirep}
\usepackage[utf8]{inputenc}
\usepackage[T1]{fontenc}
\usepackage{amsmath}
\usepackage{siunitx}
\usepackage{graphicx}
\usepackage{comment}
\usepackage{mathrsfs}
\usepackage{multirow}
\usepackage{tikz}
\usetikzlibrary{shapes,arrows,snakes,shadows.blur}

\title{\textcolor{black}{Discriminating chaotic and stochastic time series using permutation entropy and artificial neural networks}}

\author[1]{B. R. R. Boaretto}
\author[1]{R. C. Budzinski}
\author[1]{K. L. Rossi}
\author[1]{T. L. Prado}
\author[1]{S. R. Lopes}
\author[2]{C. Masoller}
\affil[1]{Department of Physics, Universidade Federal do Paran\'a, 81531-980 Curitiba, Brazil}
\affil[2]{Department of Physics, Universitat Politecnica de Catalunya, 08222, Terrassa, Barcelona, Spain}

\begin{abstract}
Extracting relevant properties of empirical signals generated by nonlinear, stochastic, and high-dimensional systems is a challenge of complex systems research. Open questions are how to differentiate chaotic signals from stochastic ones, and how to quantify nonlinear and/or high-order temporal correlations. Here we propose a new technique to reliably address both problems. Our approach follows two steps: first, we train an artificial neural network (ANN) with flicker (colored) noise to predict the value of the parameter, $\alpha$, that determines the strength of the correlation of the noise. To predict $\alpha$ the ANN input features are a set of probabilities that are extracted from the time series by using symbolic ordinal analysis. Then, we input to the trained ANN the probabilities extracted from the time series of interest, and analyze the ANN output. We find that the $\alpha$ value returned by the ANN is informative of the temporal correlations present in the time series. To distinguish between stochastic and chaotic signals, we exploit the fact that the difference between the permutation entropy (PE) of a given time series and the PE of flicker noise with the same $\alpha$ parameter is small when the time series is stochastic, but it is large when the time series is chaotic. We validate our technique by analysing synthetic and empirical time series whose nature is well established. We also demonstrate the robustness of our approach with respect to the length of the time series and to the level of noise. We expect that our algorithm, which is freely available, will be very useful to the community.  
\end{abstract}

\begin{document}

\flushbottom
\maketitle
\thispagestyle{empty}

\section*{Introduction} \label{sec:int}
Chaotic and stochastic systems have been extensively studied and the fundamental difference between them is well known: in a chaotic system an initial condition always leads to the same final state, following a fixed rule, while in a stochastic system, an initial condition leads to a variety of possible final states, drawn from a probability distribution \cite{ott2002chaos}. However, the signals generated by chaotic and stochastic systems are not always easy to distinguish and many methods have been proposed to differentiate chaotic and stochastic time series \cite{ikeguchi1997difference,rosso2007distinguishing,lacasa_2010,zunino2012distinguishing, ravetti2014distinguishing,kulp2014discriminating, quintero_njp_2015,toker2020simple,lopes2020parameter}.
A related important problem is how to appropriately quantify the strength and length of the temporal correlations present in a time series \cite{simonsen1998determination,weron2002estimating,carbone2007algorithm,witt_2013}.{\textcolor{black} {The performance of these methods varies with the characteristics of the time series. As far as we know, no method works well with all data types, because methods have different limitations, in terms of the length of the time series, the level of noise, the stationarity or seasonality of the underlying process, the presence of linear or nonlinear correlations, etc. Moreover, any time series analysis method will return, at least, one number. Therefore, to obtain interpretable results, the values obtained from the analysis of the time series of interest need to be compared with those obtained from other ``reference'' time series, where we have previous knowledge of the underlying system that generates the data. Here we use as ``reference'' model a well-known stochastic process: flicker noise (FN).}}

A FN time series is characterized by a power spectrum $P(f) \propto 1/f^\alpha$, with $\alpha$ being a quantifier of the correlations present in the signal \cite{beran2016long}. Flicker noise has been extensively studied in diverse areas such as electronics \cite{voss1976flicker,hooge1981experimental}, biology \cite{peng1992long,peng1993long}, physics \cite{bak1987self,silva2019correlated}, economy \cite{granger1996varieties,mandelbrot1997variation}, meteorology \cite{koscielny1998indication}, astrophysics \cite{press1978flicker}, etc. Furthermore, related to this issue, many methods described in the literature are able to evaluate the time correlation quantification $\alpha$, such as the Hurst exponent $\mathcal{H}$ \cite{ikeguchi1997difference,simonsen1998determination,carbone2007algorithm,beran2016long,olivares2016quantifying,weron2002estimating,silva2019correlated}.

In this paper we propose a new methodology that simultaneously allows to distinguish chaotic from stochastic time series, and to quantify the strength of the correlations. Our algorithm, based on an Artificial Neural Network (ANN)~\cite{koza1996automated}, is easy to run and freely available \cite{ann_repository}. 
We first train the ANN with flicker noise to predict the value of the $\alpha$ parameter that was used to generate the noise. The input features to the ANN are probabilities extracted from the FN time series using ordinal analysis \cite{bandt2002permutation}, {\textcolor{black}{a symbolic method widely used to identify patterns and nonlinear correlations in complex time series \cite{rosso2009detecting,rosso2009detecting2,parlitz2012classifying}. Each sequence of $D$ data points (consecutive, or with a certain lag between them), is converted into a sequence of $D$ relative values (smallest to largest), which defines an ordinal pattern. Then, the frequencies of occurrence of the different patterns in the time series define the set of ordinal probabilities, which in turn allow to calculate information-theoretic measures such as the permutation entropy (PE, described in {\it Methods})}}. {\textcolor{black}{The PE has been extensively used in the literature, due to the fact that is straightforward to calculate, and it is robust to observational noise. Interdisciplinary applications have been discussed in Ref.~\cite{zanin2012permutation} and, more recently, in a Special Issue~\cite{focusissue}.}} 

After training the ANN with different FN time series, $x_s(\alpha)$, generated with different values of $\alpha$, we input to the ANN ordinal probabilities extracted from the time series of interest, $x$, and analyze the output of the ANN, $\alpha_{\mathrm{e}}$. We find that $\alpha_{\mathrm{e}}$ is informative of the temporal correlations present in the time series $x$. Moreover, by comparing the PE values of $x$ and of $x_s(\alpha_{\mathrm{e}})$ (a FN time series generated with the value of $\alpha$ returned by the ANN), we can differentiate between chaotic and stochastic signals: the PE values of $x$ and $x_s$ are similar when $x$ is mainly stochastic, but they differ when $x$ is mainly deterministic. Therefore, the difference of the two PE values serves as a quantifier to distinguish between chaotic and stochastic signals. We use several datasets to validate this approach. We also analyze its robustness with respect to the length of the time series and noise contamination.

\begin{figure}[t] 
    \centering
    \includegraphics[width=0.95\columnwidth]{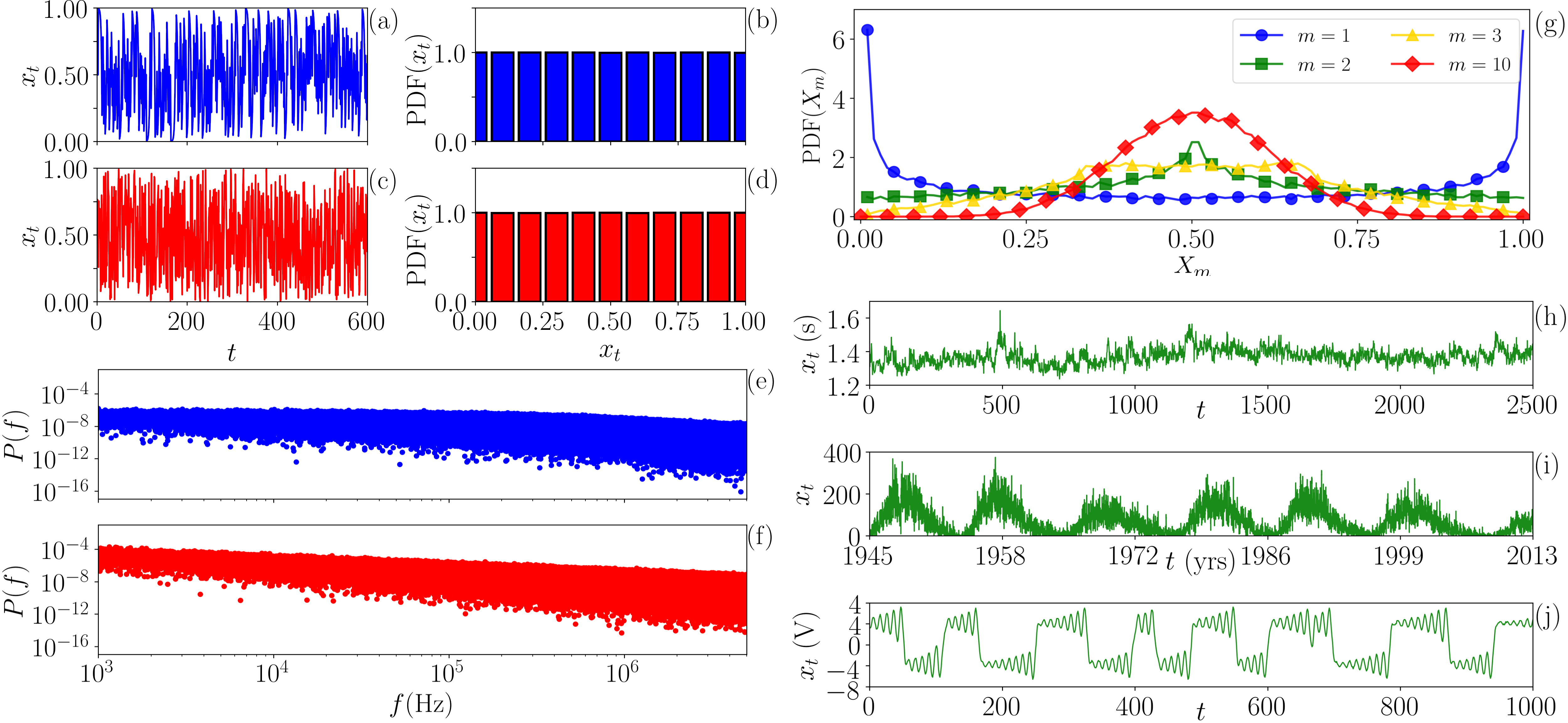}
    \caption{\textbf{Examples of time series analyzed, their probability density functions (PDFs) and power spectral densities (PSDs).} (a),(b) Time series generated by iteration of the $\beta x$ map [Eq. (\ref{eq:betax}) with $\beta=2$] and its PDF. (c),(d) Uniformly distributed white noise and its PDF. We see that the PDF of the  deterministic map is identical to the PDF of the noise. (e) and (f) PSD of the Schuster map, Eq. (\ref{eq:schuster}) with parameter $z=1.5$, and of a Flicker noise with $\alpha=1$. We note that the PSD of the Schuster map has a long decay that is very similar to a $1/f^\alpha$ decay of the noise. (g) PDF of $m$ summed logistic maps (Eq. \ref{eq:logistic} with $r=4$), which approaches a Gaussian as $m$ increases. (h-k) Examples of empirical time series analyzed: (h) a stride-to-stride of an adult walking in a slow velocity, interpreted as an stochastic process; (i) daily number of sunspots as a function of time (in years), where its  fluctuations are interpreted as stochastic; (k) voltage across the capacitor of an inductor-less Chua electronic circuit, whose oscillations are known to be chaotic.}
    \label{fig:datasets}
\end{figure}

\begin{figure}[tb!] 
   \centering
\definecolor{blue(pigment)}{rgb}{0.51, 0.56, 0.59}
\begin{tikzpicture}[scale=.9]


\draw [darkgray,->,very thick] (-4.0,0) .. controls (-4.0,-1) .. (-3.0,-1);

\draw [darkgray,->,very thick] (-1.6,-1) .. controls (-1.6,-2) .. (-0.6,-2);

\draw [darkgray,->,very thick] (0,-2) .. controls (0,-3) .. (1,-3);

\draw [darkgray,->,very thick] (3,-3) .. controls (3,-4) .. (4,-4);

\node[fill={blue(pigment)},text=white,align=left,
anchor=base
,rounded corners,blur shadow={shadow blur steps=5}] at (-1,0) {1. Extract the probabilities  of the time series.} ;

\node[fill={blue(pigment)},text=white,align=left,
anchor=base
,rounded corners,blur shadow={shadow blur steps=5}] at (2.1,-1) {2. Input them to the ANN; obtain the correlation coefficient $\alpha_\mathrm e$.} ;

\node[fill={blue(pigment)},text=white,align=left,
anchor=base
,rounded corners,blur shadow={shadow blur steps=5}] at (3.8,-2) {3. Generate a time series of Flicker noise with $\alpha=\alpha_\mathrm e$.};

\node[fill={blue(pigment)},text=white,align=left,
anchor=base
,rounded corners,blur shadow={shadow blur steps=5}] at (6.8,-3) {4. Compare the permutation entropies of the time series and of the noise.};

\node[fill={blue(pigment)},text=white,align=left,
anchor=base
,rounded corners,blur shadow={shadow blur steps=5}] at (8.1,-4) {5. Classify the time series as chaotic or stochastic.};
    \end{tikzpicture}
    \caption{\textbf{Schematic representation of the methodology.} We compute the probabilities of the ordinal patterns and then use them as input features to the ANN. The ANN returns the temporal correlation coefficient $\alpha_\mathrm e$. Then we compare the permutation entropy of the analyzed time series with the permutation entropy of a FN time series generated with the same $\alpha_\mathrm e$ value. Based on this comparison, we use Eq. (\ref{delta_h}) to classify the time series as chaotic or stochastic.}
    \label{fig:method}
\end{figure}
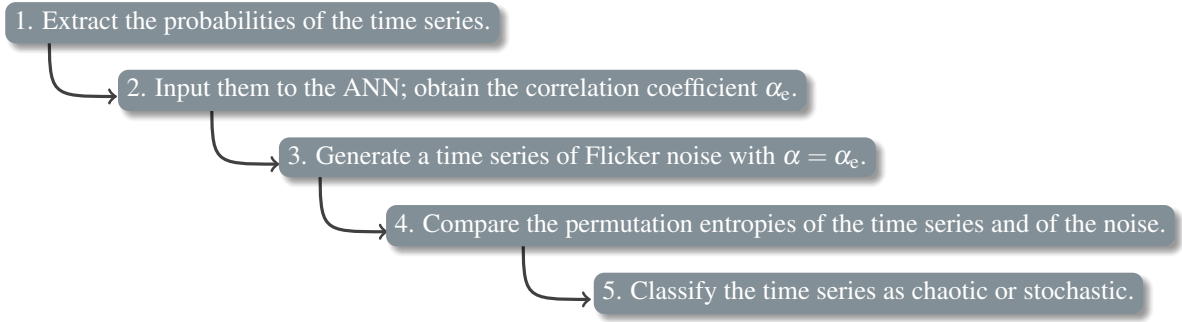

This paper is organized as follows. In the main text we present the results of the analysis of synthetic and empirical time series, which are described in section {\it Data sets}. Typical examples of the time series analyzed are presented in Fig.~\ref{fig:datasets}. In  section {\it Methods} we describe the ordinal method and the implementation of the algorithm, schematically represented in Fig.~\ref{fig:method}.

\section*{Results}\label{sec:results}
\subsection*{Analysis of synthetic datasets}

\begin{figure}[tb!] 
    \centering
    \includegraphics[width=.9\columnwidth]{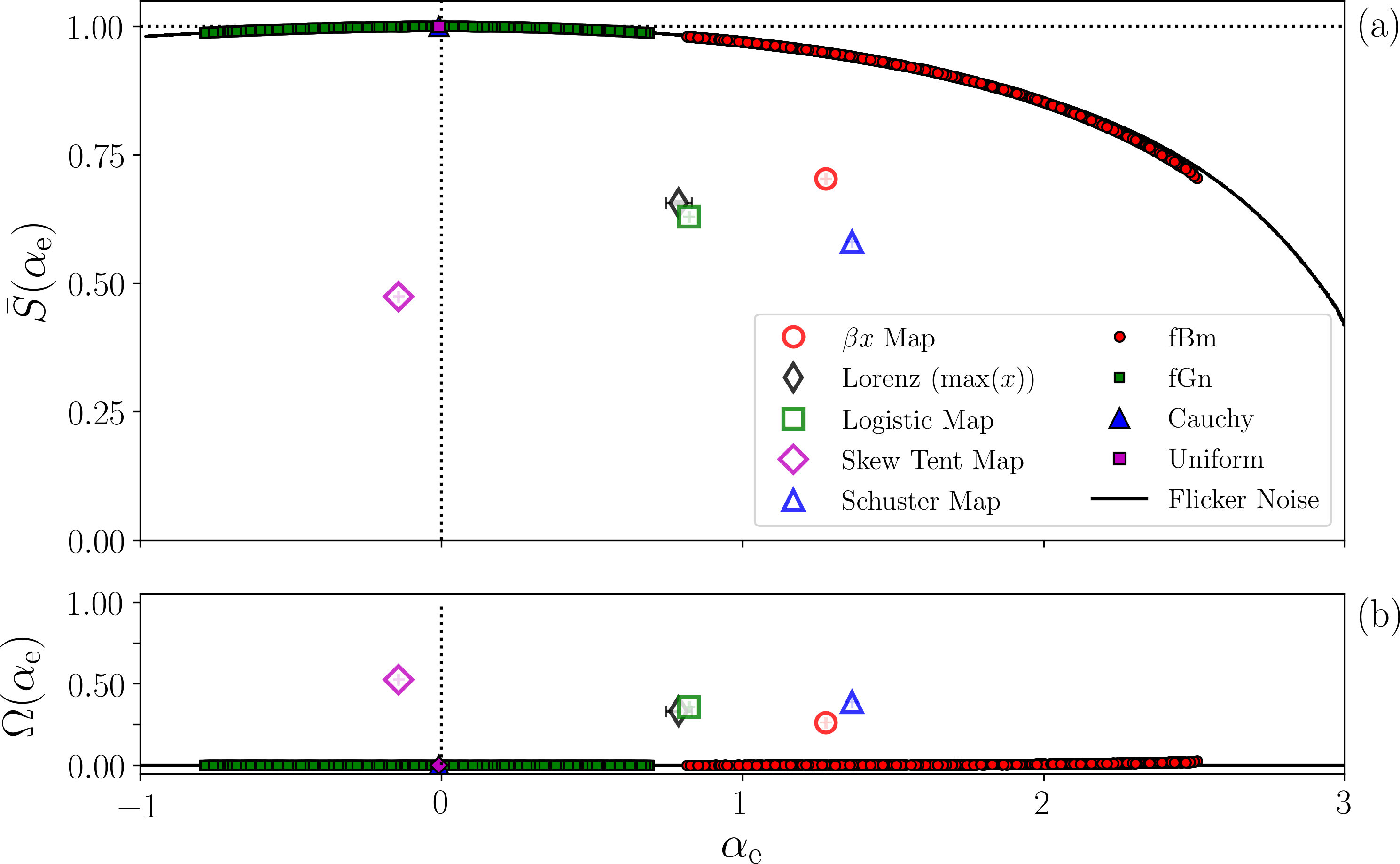}
    \caption{\textbf{Temporal correlation and distinction of chaotic and stochastic synthetic signals.} Panel (a) shows the normalized permutation entropy $\bar S$ as a function of $\alpha_{\mathrm{e}}$, evaluated through the ANN, for different time-series (stochastic and chaotic). The black solid line represents FN signals, which are used for training and testing the ANN. Filled symbols represent different stochastic signals, which are very close to the FN curve. Empty symbols represent chaotic signals, which are far away. The distance to the FN curve is measured by $\Omega$ [Eq.~(\ref{delta_h})] and it is shown in panel (b): higher values of $\Omega$ indicate chaotic time series, while lower, vanishing values indicate stochastic ones.}
    \label{fig:det_1}
\end{figure}

The main result is depicted in Fig. \ref{fig:det_1}. Panel (a) shows the normalized permutation entropy $\bar S(\alpha_\mathrm{e})$ (Eq. (\ref{eq:normalized_permutation_entropy})) vs. the time-correlation coefficient $\alpha_{\mathrm{e}}$. The filled (empty) symbols correspond to different types of stochastic (chaotic) time series, and the solid black line corresponds to FN time series generated with $\alpha \in [-1,3]$, which is accurately evaluated by the ANN. For $\alpha_\mathrm{e} = 0$, FN has a uniform power spectrum, characteristic of an uncorrelated signal (white noise), with equal ordinal probabilities $\mathcal{P}(i) \approx 1/D!$ and, hence, $\bar S = 1$. Otherwise, for $\alpha \ne 0$, some ordinal patterns occur in the time series more often than others, and the ordinal probabilities are not all equal, which decreases the permutation entropy. These results are consistent with those that have been obtained by using different methodologies \cite{kulp2014discriminating, olivares2016quantifying, lopes2020parameter}. 

In Fig.~\ref{fig:det_1} we note that fBm signals have larger time-correlation ($\alpha_\mathrm e$ closer to $2$, a classic Brownian motion) than the other three stochastic systems $\alpha_\mathrm e \approx 0$. However, their permutation entropies are very close to those of the FN signals.
The key observation is that stochastic time series all fall close to the FN curve, while chaotic ones do not, namely, $\beta x$ map, Lorenz system, logistic map, skew tent map, and Schuster map. The distance to the FN curve thus serves to distinguish stochastic and chaotic time series. This is quantified by
\begin{equation}
    \Omega(\alpha_\mathrm e) =  \frac{|\bar S_\mathrm{fn}(\alpha_{\mathrm{e}}) - \bar S|}{\bar S_{\mathrm{fn}}(\alpha_{\mathrm{e}})},
    \label{delta_h}
\end{equation}
where $\bar S$ is the permutation entropy of the analyzed time series and $\bar S_{\mathrm{fn}}(\alpha_{\mathrm{e}})$ is the PE of a flicker noise time series generated with the value of $\alpha$ returned by the ANN, $\alpha_{\mathrm{e}}$. The results are presented in Fig.~\ref{fig:det_1}, panel (b), where we see that stochastic signals have $\Omega \approx 0$, and deterministic signals have $\Omega > 0$. To summarize this finding, Table \ref{tab:table_1} depicts $\alpha_\mathrm e$ and $\Omega$ for ten representative systems.

\begin{table}[tb] 
\setlength{\tabcolsep}{7pt}
    \centering
    \begin{tabular}{l c c }
            \hline \hline
           Stochastic process $\;\;\;\;\;\;\;\;$ & $\alpha_\mathrm e$ & $\Omega (\alpha_\mathrm e)$ \\ 
           \hline 
         FN ($\alpha = 0$) & $-0.008 \pm 0.013$ & $0.00006 \pm 0.00005$  \\
         fBm ($\mathcal H = 0.5$) & $\;\;\;1.74 \pm 0.01$ & $0.005 \pm 0.001$  \\
           fGn ($\mathcal H = 0.5$) & $-0.009 \pm 0.014$ & $0.00006 \pm 0.00004$ \\
           Cauchy & $-0.008 \pm 0.004$ & $0.00007 \pm 0.00005$   \\
           Uniform & $-0.008 \pm 0.003$ & $0.00007 \pm 0.00006$  \\
         \hline \hline
              Chaotic systems & $\alpha_\mathrm e$ & $\Omega (\alpha_\mathrm e)$ \\ 
           \hline 
         $\beta x$ map & $\;\;\;1.277 \pm 0.002$ & $0.2612 \pm 0.0002$ \\
          Lorenz system ($\mathrm{max}(x)$) & $\;\;\;0.79  \pm 0.04$ & $0.33 \pm 0.004$ \\
            Logistic map & $\;\;\;0.823 \pm 0.003$ & $0.3585 \pm 0.0001$ \\
           Schuster map & $\;\;\;1.364 \pm 0.002$ & $0.3855 \pm 0.0004$ \\
           Skew Tent map & $-0.142 \pm 0.002$ & $0.5256 \pm  0.0004$ \\ \hline \hline
    \end{tabular}
    \caption{Results obtained from stochastic and deterministic time series: mean and standard deviation of the $\alpha_\mathrm e$ parameter and of $\Omega(\alpha_\mathrm e)$ (Eq.~\ref{delta_h}), calculated from $1000$ time series generated with different initial conditions and noise seeds.}
    \label{tab:table_1}
\end{table}

Next, we study the applicability of our methodology to noise-contaminated signals. We analyze the signal 
\begin{equation}
    Z_t = (1-\eta)X_t + \eta Y_t, \;\;\; t = 1,\cdots,N,
    \label{eq: noise}
\end{equation}
where $X_t$ is a deterministic (chaotic) signal ``contaminated'' by a uniform white noise, $Y_{t}$, and $\eta \in [0,1]$ controls the stochastic component of $Z_t$. For $\eta = 0 \, (1)$ the signal is fully deterministic (fully stochastic).

\begin{figure}[tb!] 
    \centering
    \includegraphics[width=\columnwidth]{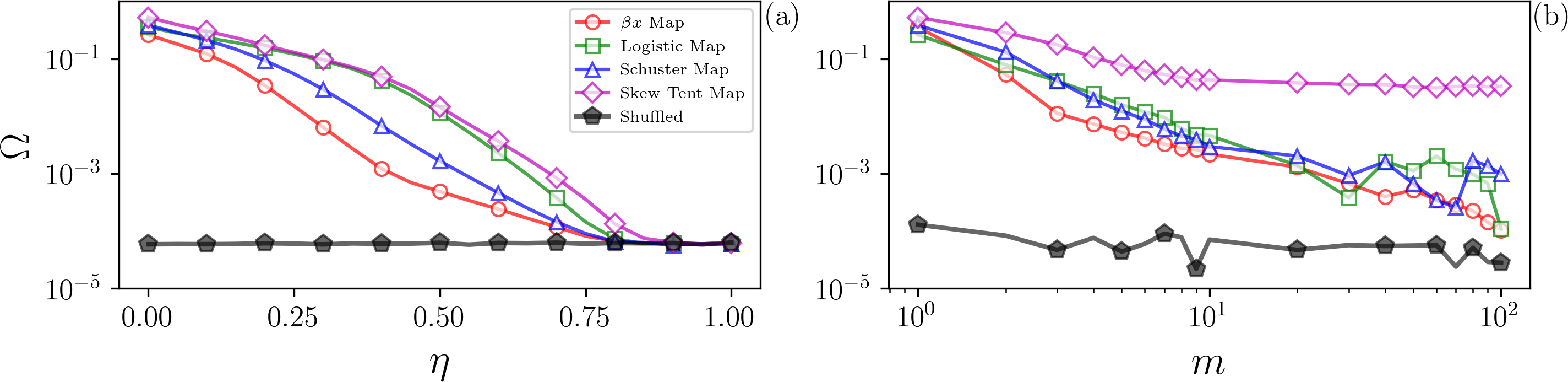}
    \caption{\textbf{Limits for identifying determinism.} Panel (a) shows the influence of noise contamination: the quantifier $\Omega$, Eq. (\ref{delta_h}) is plotted as a function of the level of noise, $\eta$, in Eq. (\ref{eq: noise}). We see that as $\eta$ increases, $\Omega$ decreases, but it remains, for high values of $\eta$, different from the value obtained from shuffled data (black pentagons). Therefore, small values of $\Omega$ can still reveal determinism in noise-contaminated signals. Panel (b) shows the effect of adding several independent chaotic signals: $\Omega$ is plotted as a function of the number $m$ of signals added. We see that as $m$ increases, $\Omega$ decreases, indicating that the deterministic nature of the signal can no longer be detected. }
    \label{fig:det_2}
\end{figure}

Figure~\ref{fig:det_2}(a) shows $\Omega$ as a function of $\eta$ for different chaotic signals. As expected, for $\eta=0$, $\Omega$ is high, but as $\eta$ grows, the level of stochasticity increases and $\Omega$ decreases. At $\eta > 0.5$, the signal is strongly stochastic, as reflected by $\Omega \approx 0.0$. For comparison, in Fig. \ref{fig:det_2}(a) we also present results obtained by shuffling a chaotic time series. As expected, $\Omega \approx 0$ for all $\eta$ because  temporal correlations are destroyed by the shuffling process.

We expect that the addition of a sufficiently large number of independent chaotic signals gives a signal that is indistinguishable from a fully stochastic one. This is verified in Fig. \ref{fig:det_2}(b), where the horizontal axis represents the number, $m$, of independent chaotic signals added. Here a high value of $\Omega$ is observed for $m = 1$ (a single chaotic signal), but as $m$ increases $\Omega \rightarrow 0$ since the chaotic nature of added signals is no longer captured (examples of the PDFs of the time series obtained from the addition of $m$ Logistic maps were presented in Fig. \ref{fig:datasets}(g), where we can observe a clear evolution towards a Gaussian shape).

\begin{figure}[tb!] 
    \centering
    \includegraphics[width=0.95\columnwidth]{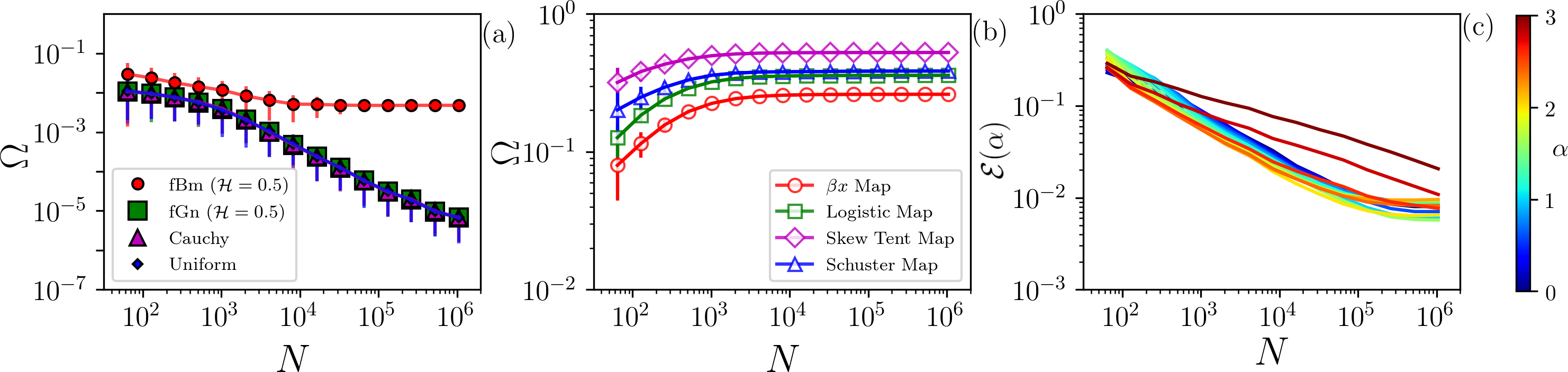}
    \caption{\textbf{Robustness with respect to the time series length.} We evaluate $\Omega$ as a function of the time series length $N$ for stochastic signals (panel (a)) and chaotic ones (panel (b)). In (a) we observe that even for $N < 100$, $\Omega < 0.1$, which confirms the stochastic nature of the signal. In (b), even for $N < 100$, $\Omega > 0.1$ (except for $\beta x$ map). Also, $\Omega > 0.2$ for all cases for $N \geq 1000$, which classifies the signals as chaotic even with only 1000 datapoints. The error bars are the standard deviation over $1000$ of samples. (c) Mean absolute error, $\mathcal{E}$, between the output of the ANN ($\alpha_{\mathrm{e}}$), and the parameter $\alpha$ used to generate the time series of flicker noise (depicted in color-code) as a function of the length of the time series, $N$.}
    \label{fig:fig_det_3}
\end{figure}

To further explore the robustness of our methodology, we investigate the role of the length $N$ of the analyzed time series in the evaluation of the $\Omega$ quantifier (Eq. (\ref{delta_h})). Figure \ref{fig:fig_det_3} shows $\Omega$ as a function of $N$, where panel (a) depicts stochastic signals, and (b) chaotic ones. We see that even for $N<10^{2}$, for all stochastic signals in panel (a) $\Omega < 0.1$, which indicates that we can identify the stochastic nature of short signals. 
For the chaotic signals in panel (b), for $N>10^2$ $\Omega > 0.1$ (except for $\beta x$ map), and for $N \geq 10^{3}$, $\Omega > 0.2$ for all signals, which demonstrates that our method can also detect determinism in short signals. 

As discussed in Sec. {\it {Methods}}, the ANN was trained with flicker noise signals with $2^{20}$ data points. However, it is interesting to analyze how much data the trained ANN needs, in order to correctly predict the $\alpha$ value of a flicker noise time series. To address this point, we generate $L=1\,000$ FN time series and analyze the error of the ANN output, $\alpha_{\mathrm{e}}$, as a function of the length of the time series, $N$, and of the value of $\alpha$ used to generate the time series. The results are presented in panel (c) of Fig.~\ref{fig:fig_det_3} that displays the mean absolute error, $\mathcal{E} =\frac{1}{L} \sum_ {\ell=1}^ L|\alpha_{\mathrm e} - \alpha|$. We see that as $N$ increases, $\mathcal{E}$ decreases. The error depends on both, $\alpha$ and $N$, and tends to be larger for high $\alpha$ due to non-stationarity and finite time sampling \cite{lopes2020parameter}. For FN time series longer than $10^{4}$ datapoints, the ANN returns a very accurate value of $\alpha$. 

\subsection*{Analysis of empirical time series}

\begin{figure}[tb!] 
    \centering
    \includegraphics[width=0.95\columnwidth]{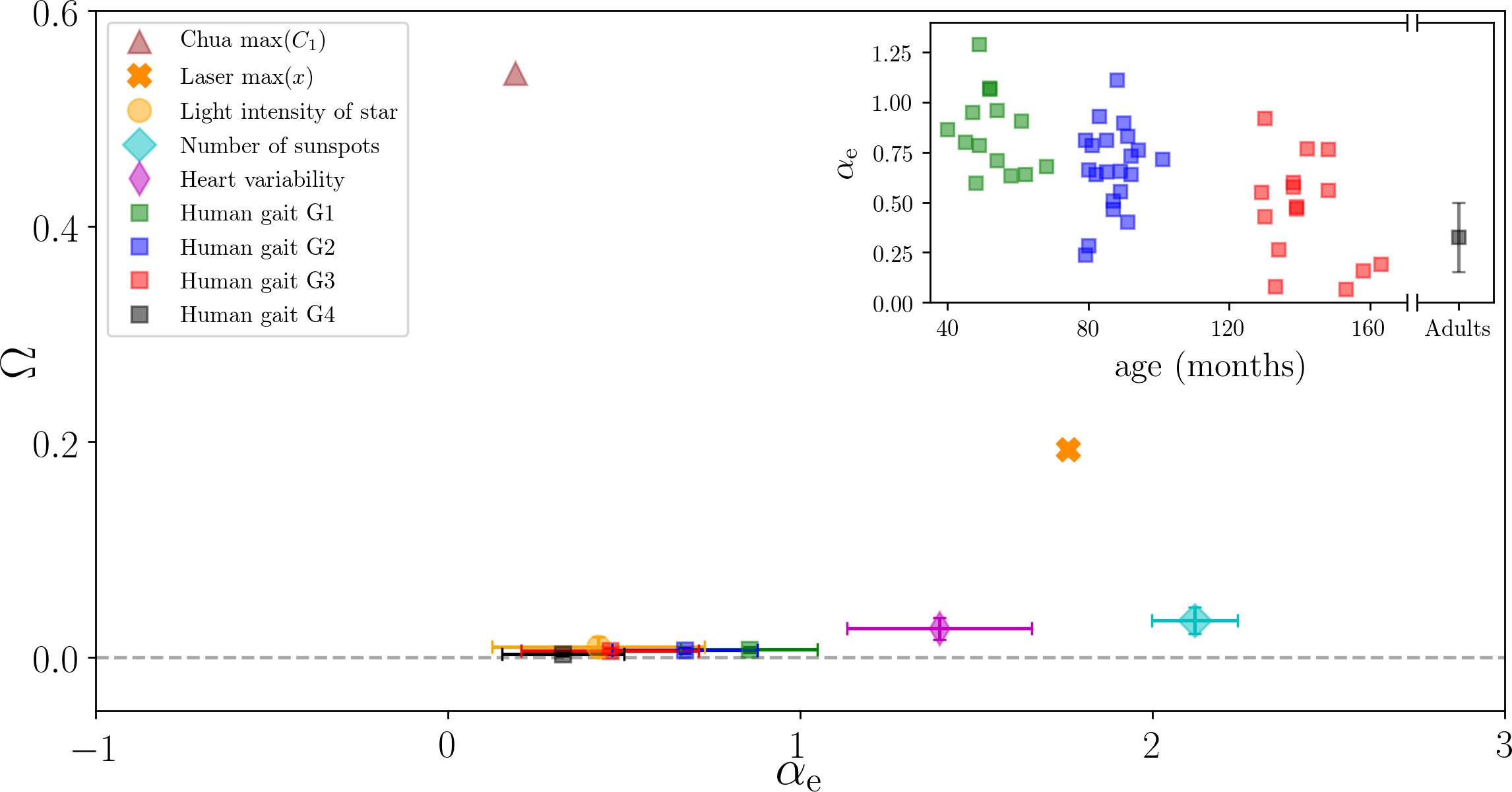}
    \caption{\textbf{Analysis of empirical time series.} Results obtained from each time series are presented in the plane ($\alpha_{\mathrm{e}}$, $\Omega$). Deterministic signals are the Chua circuit data (brown triangle) and the laser data (orange `X' marker) that have $\Omega > 0.0$. The other signals [the light intensity of a star (yellow dot), the number of sunspots (cyan diamond), the heart variability of healthy subjects (magenta thin diamond), and the different groups of human gait dynamics (green, blue, red, and black squares)] are stochastic and have $\Omega \approx 0$. For the human gait, the inset depicts the $\alpha_\mathrm e$ as a function of the age of each subject. Consistent with \cite{hausdorff1999maturation}, the correlation coefficient $\alpha_\mathrm e$ decrease with the ages.
    }
    \label{fig:experimental}
\end{figure}

Here we present the analysis of time series recorded under very different experimental conditions, as described in section {\it Data sets}. 
Figure \ref{fig:experimental} displays the results in the plane ($\alpha_{\mathrm{e}}$, $\Omega$). The $\Omega$ values obtained for the Chua circuit data and for the laser data confirm their chaotic nature \cite{chua1994chua,gershenfeld1993future} ($\Omega \approx 0.55$ and $\Omega \approx 0.20$ respectively). For the star light intensity $\Omega \approx 0$, confirming the stochastic nature of the signal~\cite{weigend2018time}. For the number of sunspots, which is a well-known long-memory noisy time series, $\Omega \approx 0$. In this case the value of $\alpha$ obtained ($\alpha \approx 2$) confirms the results of Singh \textit{et al.} \cite{singh2017early} where a Hurst exponent close to 1 was found. Regarding the five time series of RR-intervals of healthy subjects, our algorithm identifies stochasticity ($\Omega \approx 0$) in all of them, which is consistent with findings of Ref. ~\cite{toker2020simple}. 

The last empirical set analyzed reveals the nature of the dynamics of human gait: regardless of the age of the subjects, $\Omega \approx 0$ confirming the stochastic behavior discussed in \cite{hausdorff1996fractal}. In the inset we show that the $\alpha_\mathrm e$ value returned by the ANN decreases with the age, which is also in line with the results presented in \cite{hausdorff1999maturation}, obtained with Detrended Fluctuation Analysis (see Fig. 6 of Ref.~\cite{hausdorff1999maturation}). The authors interpret this variation as due to an age-related change in stride-to-stride dynamics, where the gait dynamics of young adults (healthy) appears to fluctuate randomly, but with less time-correlation in comparison to young children \cite{hausdorff1999maturation}.

\section*{Discussion}\label{sec:discussions}

We have proposed a new time series analysis technique that allows to differentiate stochastic from chaotic signals, and also, to quantify temporal correlations. We have demonstrated the methodology by using synthetic and empirical signals.

Our method is based on locating a time series in a two dimensional plane determined by the permutation entropy and the value of a temporal correlation coefficient, $\alpha_e$, returned by a machine learning algorithm. In this plane, stochastic signals are very close to a curve defined by Flicker noise, while chaotic signals are located far from this curve. We have used this fact to define a quantifier, $\Omega$, that is the distance to the FN curve. $\Omega$ serves to distinguish stochastic and chaotic time series, and it can be used to analyze time series, even when they are very short (with time series of 100 datapoints we found that $\Omega<0.1$ or $\Omega>0.1$, if the time series is stochastic or chaotic respectively, Fig.~\ref{fig:fig_det_3}). We also found that small values of $\Omega$ can be used to identify underlying determinism in noise-contaminated signals, and in signals that result from the addition of a number of independent chaotic signals (Fig.~\ref{fig:det_2}). We have also used our algorithm to analyze six empirical datasets, and obtained results that are consistent with prior knowledge of the data (Fig.~\ref{fig:experimental}). {\textcolor{black}{Taken together, these results show that the proposed methodology allows answering the questions of how to quantify stochasticity and temporal correlations in a time series.}} 

Our algorithm is fast, easy-to-use, and freely available~\cite{ann_repository}. Thus, we believe that it will be a valuable tool for the scientific community working on time series analysis. {\textcolor{black}{Existing methods have limitations in terms of the characteristics of the data (length of the time series, level of noise, etc.)}}. A limitation of our algorithm lies in the analysis of noise-contaminated periodic signals, because their temporal structure may not be distinguished from the temporal structure of a fully stochastic signal with a large $\alpha$ value. Future work will be directed at trying to overcome this limitation. {\textcolor{black}{Here, for a ``proof-of-concept'' demonstration, we have used a well-known machine learning algorithm (a feed-forward ANN), a rather simple training procedure, and a popular entropy measure (the permutation entropy). We have not tried to optimize the performance of the algorithm. We expect that different machine learning algorithms, training procedures, and entropy measures can give different performances, depending on the characteristics of the data analyzed. Therefore, the methodology proposed here has a high degree of flexibility, which can allow to optimize performance for the analysis of particular types of data.}}

\section*{Methods}

\subsection*{Ordinal analysis and permutation entropy}

\begin{figure}
    \centering
    \includegraphics[width=.9\columnwidth]{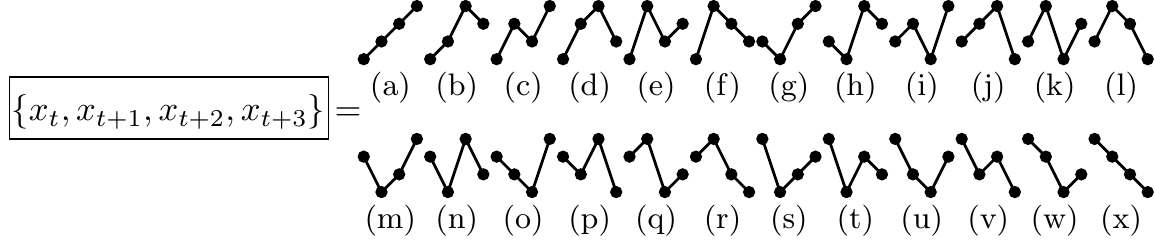}
    \caption{\textcolor{black}{Schematic illustration of 24 ordinal patterns that can be defined from $D=4$ consecutive data values in a time series.}}
    \label{fig:ordinal_patterns}
\end{figure}

Ordinal analysis allows the identification of patterns and nonlinear correlations in complex time series \cite{bandt2002permutation}. For each sequence of $D$ data points in the time-series (consecutive, or with a certain lag between them), their values are replaced by their relative amplitudes, ordered from $0$ to $D-1$. {\textcolor{black}{For instance, a sequence $\{0,5,10,13\}$ in the time series transforms into the ordinal pattern ``0123'', while $\{0,13,5,10\}$ transforms into ``0312''. As an example, Fig.~\ref{fig:ordinal_patterns} shows the ordinal patterns formed with $D=4$ consecutive values.}}. 

We evaluate the frequency of occurrence of each word, defined as the ordinal probability $\mathcal{P}(i)$ with $\sum_{i=1}^{D!}\mathcal{P}(i)=1$, where $i$ represents each possible word. Then, we evaluate the Shannon entropy, known as permutation entropy \cite{bandt2002permutation}:
\begin{equation}
    S(D) = - {\sum_{i=1}^{D!}\mathcal{P}(i)\ln{(\mathcal{P}(i)})}.
    \label{eq:permuation_entropy}
\end{equation}
The permutation entropy varies from $S(D)=0$ if the $j$-th state $\mathcal P(j)=1$ (while $\mathcal P(i)=0$ $\forall\; i\ne j$) to $S(D)=\ln({D!})$ if $\mathcal{P}(i)=1/D!$ $\forall\; i$. The normalized permutation entropy used in this work is given by:
\begin{equation}
    \bar S(D) = \frac{S(D)}{\ln D!}.
    \label{eq:normalized_permutation_entropy}
\end{equation}

\textcolor{black}{In this work, to calculate the ordinal patterns, we have used the algorithm proposed by Parlitz and coworkers~\cite{parlitz2012classifying}. We have used $D =  6$ and no lag, i.e., $D-1$ values overlap in the definition of two consecutive ordinal patterns. Therefore, we use as features to the ANN (see below) the $D!=720$ probabilities of the ordinal patterns. For a robust estimation of these probabilities, a time series of length $N>>D!$ is needed. However, as we show in Fig.~\ref{fig:fig_det_3}, the algorithm returns meaningful values even for time series that are much shorter.}

\subsection*{Artificial Neural Network} 

\textcolor{black}{Deep learning is part of a broader family of machine learning methods based on artificial neural networks (ANNs) \cite{wiki}. In this work, we use the deep learning framework \textit{Keras} \cite{kerasio} to compile and train an ANN. Since we want to regress the information of the features into a real value (classical scalar regression problem \cite{chollet2017deep}) an appropriate option is a feed-forward ANN. The ANN is trained to evaluate the time correlation coefficient considering as features the $720$ probabilities of the ordinal patterns. We connect the input layer to a single dense layer with $64$ output units connected to a final layer, regressing all the information of the inputs into a real number. Other combinations were tested with different numbers of units ($16,\,512,\,1024$) and layers. We found that a single layer with 64 units was sufficient to accurately predict the $\alpha$ value. The ANN parameters and the compilation setup are given in Table \ref{table_feed_forward}. As explained in the discussion we have used the feed-forward ANN as a simple option for a “proof-of-concept” demonstration. Other deep learning/machine learning methods or a different compilation setup may give better results depending on the type of data that is analyzed.}

\begin{table}[htb!]
\centering
\caption{\textcolor{black}{Compilation setup and parameters of the feed forward ANN.}}
\vspace*{0.0cm}
\begin{tabular}{l l l l}
    \hline \hline
    Compilation setup &\hspace{3.8cm} & & \\
    \hline 

    Optimizer &  & & Adam \\
    Loss function & & & Mean square error  \\
    Metrics & & & Mean absolute error
\end{tabular}
\begin{tabular}{l l r l}
    \hline \hline
    Trainable Parameters &  &  & \\
\hline 
    Layer (type) & Output Shape & Param \# & activation\\
\hline 
Dense \# 1 & $64$ & 46144 & `relu' \\
Dense \# 2 & $1$ & 65 & None\\
\hline \hline
& & Total params: 46209
\end{tabular}
\label{table_feed_forward}
\end{table}

\textcolor{black}{The training stage of the ANN is performed using a dataset of $50\,000$ flicker noise time series with $N=2^{20}$ points, where the parameter $\alpha$ of each time series is randomly chosen in $-1\leq \alpha \leq 3$ (see Section {\it Datasets} for details). We separate the dataset into two sets: the training dataset ($L_\mathrm{train}=40\,000$), and the test dataset ($L_\mathrm{test}=10\,000$). To quantify the error between the output of the ANN and the target, $\alpha$, we use the \textit{mean absolute error}:
\begin{equation}
    \mathcal{E} =\frac{1}{L} \sum_ {\ell=1}^ L|\alpha_{\mathrm e,\ell} - \alpha_{\ell}|.
    \label{eq:mae}
\end{equation}
where $L$ is the number of samples. The training stage is concluded and then the parameters of the ANN are fixed. After that, we apply the ANN to the test dataset, and the error in the evaluation of $\alpha_{\mathrm{e}}$ regarding $\alpha$ is $\mathcal{E}(L_\mathrm{test}) \approx 0.01$.}

\section*{Datasets} \label{sec:datasets}

\subsection*{Stochastic systems}\label{sec:stochastic_data}

In this paper we use three types of stochastic signals: flicker noise (FN), fractional Brownian motion (fBm) and fractional Gaussian noise (fGn). Flicker noise (FN) or colored noise time series are used for the training of the Artificial Neural Network. They are generated with the open Python library \textit{colorednoise.py} \cite{colorednoisepy,timmer1995generating}. 
fBm and fGn time series are generated with the Python library \textit{fmb.py} \cite{fbmpy}. Both time series depend on a Hurst index $\mathcal H$ \cite{zunino2007characterization}. For the fBm $\mathcal H = 0.5$ corresponds to the classical Brownian motion. If $\mathcal H > 0.5$ ($\mathcal H < 0.5$) the time-series is positively (negatively) correlated. For fGn $\mathcal H = 0.5$  characterizes a white noise \cite{zunino2007characterization}; if $\mathcal H>0.5$ ($\mathcal H<0.5$) the fGn time series exhibits long-memory (short-memory). The Hurst index is related to the $\alpha$ of flicker noise: for a fBm stochastic process, $\alpha = 2\mathcal H+1$ and $1<\alpha<3$; for fGn, $\alpha = 2\mathcal H-1$ and $-1<\alpha<1$ \cite{zunino2007characterization}. 

\subsection*{Chaotic systems} 
In this paper, we analyze time series generated by five chaotic systems:

1) The generalized Bernoulli chaotic map, also known as $\beta x$ map, described by:
\begin{equation}\label{eq:betax}
    x_{t+1} = \beta x_t \; \mathrm{(mod1)},
\end{equation}
where $\beta$ controls the dynamical characteristic of the map. Throughout this paper, we use $\beta = 2$, which leads to a chaotic signal \cite{ott2002chaos}. 

2) The well-known logistic map \cite{ott2002chaos}:
\begin{equation}\label{eq:logistic}
    x_{t+1} = r x_t(1-x_t),
\end{equation}
we use $r = 4$ to obtain a chaotic signal \cite{ott2002chaos}.

3) The Schuster map \cite{schuster2006deterministic}, which exhibits intermittent signals with a power spectrum $P(f)\sim 1/f^{z}$. It is defined as:
\begin{equation}\label{eq:schuster}
    x_{t+1} = x_t + x^z _t \; \mathrm{(mod1)},
\end{equation}
where we use $z = 0.5$.

4) The skew tend map, which is defined as 
\begin{equation}\label{eq:tent}
    x_{t+1} = \begin{cases} x_t/\omega &\mbox{if } x_t \in [0,\omega], \\
(1-x_t)/(1-\omega) & \mbox{if } x_t \in (\omega,1]. \end{cases}
\end{equation}
Here, we use $\omega = 0.1847$ in order to obtain a chaotic signal \cite{rosso2007distinguishing}.

5) The also well-known Lorenz system, defined as:
\begin{eqnarray}\label{eq:lorenz}
    \frac{dx(t)}{dt} & = & \sigma (y-x),\\
    \frac{dy(t)}{dt} & = & x(R-z) -y,\\
    \frac{dz(t)}{dt} & = & xy-bz.
\end{eqnarray}
with parameters $\sigma = 16$,  $R = 45.92$, and $b = 4$, which lead to a chaotic motion \cite{wolf1985determining}. For analyze the time series of consecutive maxima of the $x$ variable. 


%
%


%
%

\subsection*{Empirical datasets}

We test our methodology with empirical datasets recorded from diverse chaotic or stochastic systems. Additional information of the datasets can be found in Table \ref{tab:table_2}. These are: 

\textbf{Dataset E-I:} Data recorded from an inductorless Chua's circuit constructed as \cite{torres2000inductorless}. The circuit was set up and the data was kindly sent to us by Vandertone Santos Machado \cite{exp_data_I}. The voltages across the capacitors depict chaotic oscillations. To detect this chaoticity we compute the maxima values of the first capacitor $C_1$.   

\textbf{Dataset E-II:} Fluctuations in a chaotic laser data approximately described by three coupled nonlinear differential equations \cite{gershenfeld1993future}. To detect the chaoticity of the laser, we compute the maxima values of the time series. The data is available in \cite{exp_data_II}.

\textbf{Dataset E-III:} Light intensity of a variable dwarf star (PG1599-035) \cite{gershenfeld1993future} with $17$ time-series (segments). These variations may be caused by an intrinsic change in emitted light (superposition of multiple independent spherical harmonics \cite{gershenfeld1993future}), or by an object partially blocking the brightness as seen from Earth. The fluctuations in the intensity of the star have been observed to result in a noisy signal \cite{gershenfeld1993future}. The data is open and freely available in \cite{exp_data_III}.

\textbf{Dataset E-IV:} Three time-series of the sunspots numbers for the period of 1976 -- 2013 \cite{singh2017early}, the daily sunspots numbers depicts a noisy "pseudo-sinusoidal" behavior. It is accepted that magnetic cycles in the Sun are generated by a solar dynamo produced through nonlinear interactions between solar plasmas and magnetic fields \cite{allen2010derivation,choudhuri2000current}. However, the fluctuations in the period in the cycles is still difficult to understand \cite{passos2008low}. This type of data has been analyzed in \cite{singh2017early}, where its stochastic fluctuations depict a Hurst exponent $\mathcal H \approx 1$, meaning the data carries memory. The data can be found at \cite{exp_data_IV_1,exp_data_IV_2,exp_data_IV_3}.

\textbf{Dataset E-V:} Five RR-interval time-series from healthy subjects. Each time series have $\sim 100\,000$ RR intervals (the signals were recorded using continuous ambulatory electrocardiograms during 24 hours). It still a debate if the heart rate variability is chaotic or stochastic \cite{toker2020simple}. While some studies suggest that heart rate variability is a stochastic process \cite{toker2020simple,baillie2009normal,zhang2009stochastic}. Much chaos-detection analysis has been identified as a chaotic signal \cite{toker2020simple,glass2009introduction}. The dataset is open and freely available in \cite{exp_data_V}.



\textbf{Dataset E-VI:} Fractal dynamics of the human gait as well as the maturation of the gait dynamics. The stride interval variability can exhibit randomly fluctuations with long-range power-law correlations, as observed in \cite{hausdorff1996fractal}. Moreover, this time-correlation tends to decrease in older children \cite{hausdorff1996fractal,hausdorff1999maturation}. The analyzed dataset is then separated into $3$ groups, related to the subjects' ages. Group No. $1$ has data for $n = 14$ subjects with 3 - to 5 - yrs old; Group No. $2$ has data for $n = 21$ subjects with 6 - to 8 - yrs old; Group No. $3$ has data for $n = 15$ subjects with 10 - to 13 - yrs old; Group No. $4$ has data for $n = 10$ subjects with 18 - to 29 - yrs old \cite{hausdorff1996fractal}. The data is open and freely available in \cite{goldberger2000PhysioBank,exp_data_VI_1,exp_data_VI_2}.

\begin{table}[htb]
\setlength{\tabcolsep}{7pt}
    \centering
    \begin{tabular}{l c c c}
            \hline \hline
        Dataset & Number of samples & Mean length $\langle N\rangle$ & Availability \\ \hline
        {\bf E-I} & $1$ & $6\,000$ & \cite{exp_data_I} \\
        {\bf E-II} & $1$ & $750$ & \cite{exp_data_II} \\
        {\bf E-III} & $17$ & $1\,600$ & \cite{exp_data_III} \\
        {\bf E-IV} & $3$ & $15\,000$ & \cite{exp_data_IV_1,exp_data_IV_2,exp_data_IV_3}  \\
         {\bf E-V} & $5$ & $100\,000$ & \cite{exp_data_V} \\
         {\bf E-VI} & $60$ & $800$ & \cite{exp_data_VI_1,exp_data_VI_2}  \\
         
         \hline \hline
    \end{tabular}
    \caption{Characteristics of empirical datasets}
    \label{tab:table_2}
\end{table}

%

\section*{Acknowledgments}   

The authors thank Vandertone Santos Machado for providing the deterministic data of a chaotic Chua circuit. B.R.R.B., R.C.B, K.L.R., T.L.P. and S.R.L acknowledge partial support of Conselho Nacional de Desenvolvimento Científico e Tecnológico, CNPq, Brazil, Grant No. 302785/2017-5, 308621/2019-0 and the Coordenação de Aper\-fei\-çoamento de Pessoal de Nível Superior, Brasil (CAPES), Finance Code 001. 
C.M. acknowledges funding by the Spanish Ministerio de Ciencia, Innovacion y Universidades (PGC2018-099443-B-I00) and the ICREA ACADEMIA program of Generalitat de Catalunya.

\end{document}